\documentclass[aps,pra,onecolumn,superscriptaddress]{revtex4-2}

\usepackage{graphicx}
\usepackage{xcolor}
\usepackage{amsmath}
\usepackage{lineno}
\begin{document}
	
\title{Excitation of $^{87}$Rb Rydberg atoms to nS and nD states (n$\leq$68) via an optical nanofiber}

\author{Alexey Vylegzhanin}\altaffiliation{these authors contributed equally}\affiliation{Light-Matter Interactions for Quantum Technologies Unit, Okinawa Institute of Science and Technology Graduate University, Onna, Okinawa 904-0495, Japan.}
\author{Dylan~J.~Brown}\altaffiliation{these authors contributed equally}\affiliation{Light-Matter Interactions for Quantum Technologies Unit, Okinawa Institute of Science and Technology Graduate University, Onna, Okinawa 904-0495, Japan.}
\author{Aswathy Raj}\affiliation{Light-Matter Interactions for Quantum Technologies Unit, Okinawa Institute of Science and Technology Graduate University, Onna, Okinawa 904-0495, Japan.}
\author{Danil~F.~Kornovan}
\affiliation{Center for Complex Quantum Systems, Department of Physics and Astronomy, Aarhus University, Ny Munkegade 120, DK-8000 Aarhus C, Denmark.}
\author{Jesse~L.~Everett}\affiliation{Light-Matter Interactions for Quantum Technologies Unit, Okinawa Institute of Science and Technology Graduate University, Onna, Okinawa 904-0495, Japan.}
\author{Etienne Brion}
\affiliation{Laboratoire Collisions Agr\'{e}gats R\'{e}activit\'{e}, UMR5589, Universit\'{e} Toulouse III Paul Sabatier, CNRS, F-31062 Toulouse Cedex 09, France.}
\author{Jacques Robert}
\affiliation{Universit\'e Paris-Saclay, CNRS, {Laboratoire de Physique des Gaz et des Plasmas}, 91405 Orsay, France.}
\author{S\'{i}le~{Nic Chormaic}}\affiliation{Light-Matter Interactions for Quantum Technologies Unit, Okinawa Institute of Science and Technology Graduate University, Onna, Okinawa 904-0495, Japan.}

\begin{abstract}
Cold Rydberg atoms are a promising platform for quantum technologies and combining them with optical waveguides has the potential to create robust quantum information devices. Here, we experimentally observe the excitation of cold rubidium atoms to a large range of Rydberg S and D states through interaction with the evanescent field of an optical nanofiber.  We  develop  a theoretical model to  account for experimental phenomena present such as the AC Stark shifts and the Casimir-Polder interaction. This work strengthens the knowledge of Rydberg atom interactions with optical nanofibers and is a critical step toward the implementation of all-fiber quantum networks and waveguide QED systems using highly excited atoms. 
\end{abstract}



\maketitle

\section{Introduction}\label{sec:introduction}
Cold Rydberg atoms are a promising platform for quantum information~\cite{demonstration_of_C_NOT_gate, quantum_information_Saffman, adams2019rydberg_quantum_tech, jaksch2000fast_gates_Rydberg, PhysRevResearch.5.013205, chew2022ultrafast} and quantum simulation~\cite{weimer2010rydberg_simulator, bharti2022ultrafast} due to the long lifetimes of the excited states and the strong dipole interaction resulting in Rydberg blockade~\cite{Urban2009}. The Rydberg blockade allows for the deterministic entanglement of qubits~\cite{Rydberg_Bell_states}, implementation of C-NOT~\cite{demonstration_of_C_NOT_gate} and C-Phase quantum gates~\cite{one_pulse_CPhase}, and the generation of single-photon emitters~\cite{ripka2018room_single_photon} and single-photon switches~\cite{baur2014single_switch}.  Typically, Rydberg experiments have been performed in free-space, often with atoms being excited to a Rydberg state in optical tweezers arrays~\cite{Rydberg_in_tweezers}, optical lattices~\cite{Rydberg_in_optical_lattices} or micron-sized vapor cells~\cite{mum_sized_cell}. However, new platforms of interest include hybrid systems such as atom-waveguide experiments~\cite{Ke:16,PhysRevA.102.063703,hollowfiber2014rydberg}, atom chips~\cite{atom_chip_excitation}, and cavity QED~\cite{cavity_Rydberg}. Compared to free-space systems, these have advantages such as low power consumption and high scalability~\cite{quantum_information_Saffman}, which are important features for creating practical quantum devices~\cite{Acín_2018}. 

Here, we present our work on Rydberg atom excitation using an optical nanofiber (ONF) integrated into a cloud of cold neutral atoms, to develop an ONF - atom hybrid  system~\cite{KP_rydberg_generation} for the purposes of developing a robust system for waveguide QED experiments~\cite{RevModPhys.95.015002}. Optical nanofibers are versatile  for many different quantum experiments~\cite{nieddu2016ONF_dev, ONF_quantum_dev} based on their ease of integration into atomic systems~\cite{nieddu2016ONF_dev}. The high intensity of the evanescent field decaying from the fiber's surface~\cite{fiber_field, Gokhroo_2022} make them ideal for manipulating and probing cold atoms~\cite{nayak_first_fiber_probe, PhysRevLett.104.203603,gouraud2015demonstration_atomic_memory,PhysRevResearch.4.023002}, with the steep gradient allowing for the study of quadrupole excitations difficult to access via free-space ~\cite{quadrupole_excitations, kien2022}. The extension of the evanescent field  far from the surface facilitates atom-light interactions many wavelengths from the fiber~\cite{Finkelstein}. In addition, ONFs are relatively easy to install into experimental setups making them a strong  candidate for compact quantum devices, all-fiber integrated quantum systems, and waveguide QED \cite{RevModPhys.95.015002,lechner2023light}. The ability to generate an ordered array of Rydberg atoms near an optical nanofiber will provide a major new direction for waveguide QED and could lead to the realization of self-ordered trains of photons through the nanofiber due to a repulsive photon-photon interaction~\cite{PhysRevResearch.4.023002}.

We experimentally excite $^{87}$Rb atoms from a magneto-optical trap (MOT) to a wide range of Rydberg states, as high as $n$ = 68 for the $n$D$_{5/2}$ state, using the evanescent field of an optical nanofiber, see Fig.~\ref{fig:experiment}(a) for a schematic of the setup. Atoms excited to the Rydberg state are lost from the MOT and we measure the remaining MOT population as an indirect measurement of the Rydberg excitation rate, see Sec.~\ref{sec:methods} and Sec.~\ref{sec:numerics} for details. 
We perform numerical calculations of the atom-surface interaction and extend our previous model~\cite{KP_rydberg_generation} using experimental parameters to better account for the full dynamics. We present the conclusions and perspectives of the work in Sec.~\ref{sec:conclusion}.

\section{Experimental details}\label{sec:setup}
\begin{figure}[!t]
    \centering
    \includegraphics[width=0.48\textwidth]{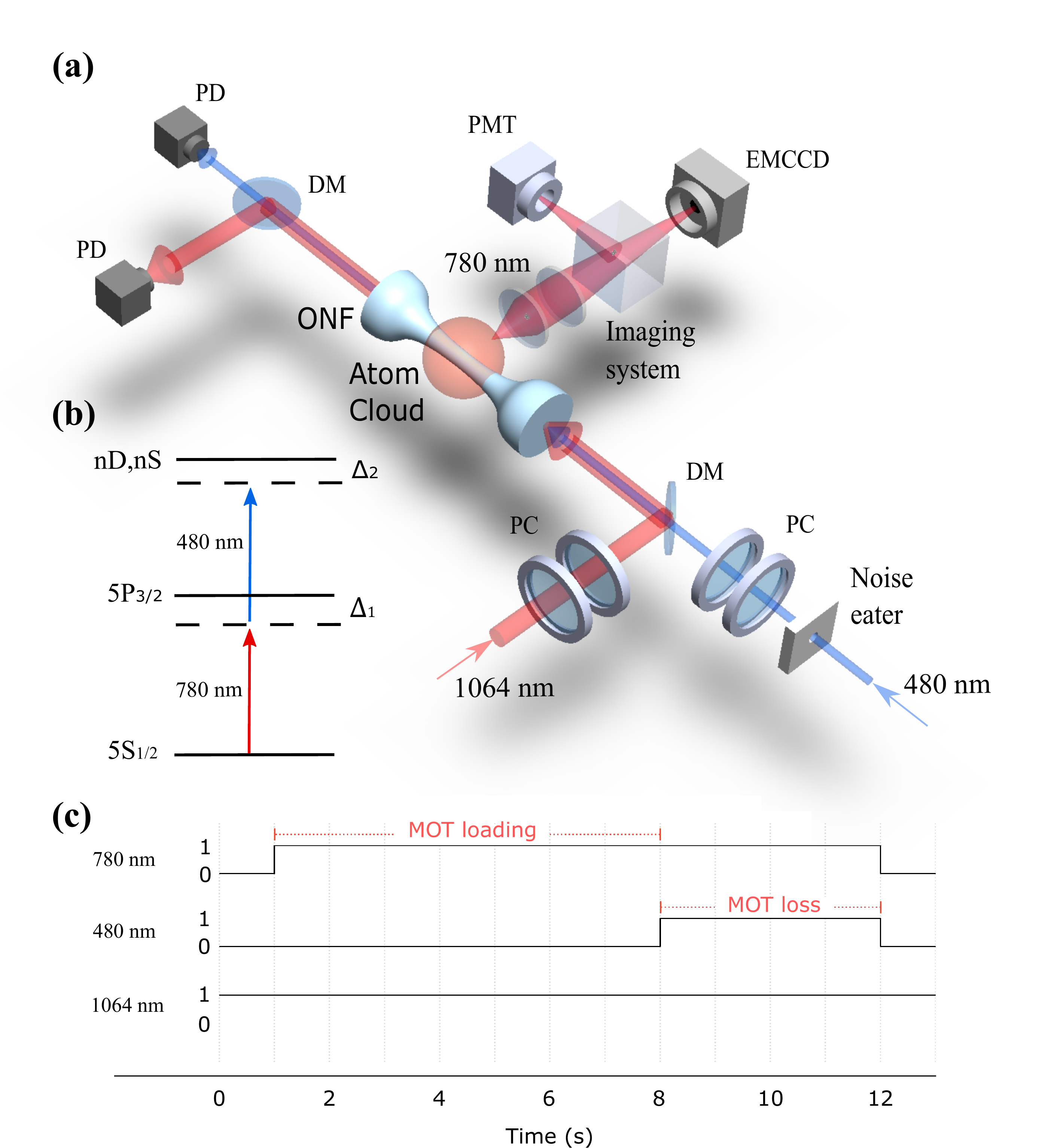}
    \caption{(a) Schematic of the experimental setup.  $^{87}$Rb atoms are trapped around the ONF  in a MOT formed from six counterpropagating 780~nm beams. The excitation to the Rydberg state is driven by 480~nm light coupled into the ONF. The polarization of the 480~nm light is set using two $\lambda$/4 waveplates to maximize the excitation rate. The MOT population is measured using a PMT by collecting the fluorescence from the trapped atoms.  ONF - optical nanofiber, PD - photodetector, DM - dichroic mirror, PC - polarization compensation, PMT - photomultiplier tube, EMCCD - electron multiplying CCD.
    (b) Energy level diagram for Rydberg excitation. The 780~nm cooling laser drives a transition between the 5S$_{1/2}$ and 5P$_{3/2}$ states, while the 480~nm laser excites atoms to the Rydberg state. The cooling and Rydberg lasers have detunings $\Delta_1$ and $\Delta_2$ from their respective transitions.
    (c) The timing sequence of the experiment. The MOT initially loads for 7 seconds.  Once the 480~nm laser is turned on, the MOT starts losing atoms. After 4 seconds, both the 480~nm and 780~nm lasers are turned off and the experiment cycle recommences. The 1064~nm laser remains at a constant power throughout the experiments.}
    \label{fig:experiment}
    \rule{0.47\textwidth}{0.3pt}
\end{figure}

The experiment is performed using a cold cloud of $^{87}$Rb atoms overlapped with an optical nanofiber with diameter $d$ $\approx$ 350~nm, see Fig.~\ref{fig:experiment}(a).    The ONF is prepared from standard 125~$\mu$m diameter fiber (SM800-5.6-125) using a H:O flame brushing technique \cite{fiber_pulling_rig}. During the tapering process, transmission through the fiber is monitored with a photodetector to ensure it remains above 99\% for 780~nm guided light. The ONF is installed in an ultrahigh vacuum (UHV) chamber, maintained at a pressure of $10^{-9}$~mbar, with the MOT centered at the waist. During the experiments, we propagate 300~$\mu$W of 1064~nm light  through the fiber to keep it hot and reduce the adsorption of $^{87}$Rb on its surface. Note we assume only  HE$_{11}$ mode propagation for all the wavelengths (480~nm, 780~nm, 1064~nm) used in the experiments.  However, the ONF diameter is  very close to the cutoff of 352~nm for the TE$_{01}$ and TM$_{01}$ guided modes at 480~nm~\cite{PhysRevA.96.023835}. 

The cold $^{87}$Rb atoms, with an average temperature of $\sim$140~$\mu$K,  are created in a standard MOT with three pairs of counterpropagating 780~nm cooling beams.  The cooling beams are stabilized to the 5S$_{1/2}(F=2)\rightarrow5\mathrm{P}_{3/2}(F=2,3)$ crossover peak and shifted with an AOM to a fixed detuning, ${\Delta_1=-14}$~MHz, from the transition to the 5P$_{3/2}(F=3)$ state.  The total power in all cooling beams is 50~mW. A pair of anti-Helmholtz coils provides a magnetic field gradient of 20~G/cm and three pairs of compensation coils allow us to overlap the atom cloud with the ONF. With no defined quantization axis, the atoms are assumed to equally occupy all Zeeman sublevels. Two EMCCD cameras (Thorlabs DCC1545M and Andor Luca R) capture images of the atom cloud. Fluorescence from the atoms is collected by a photomultiplier tube (PMT,Hamamatsu R636-10). This provides a signal that is proportional to the number of atoms trapped in the MOT.

The 480~nm light used to drive the Rydberg transitions, see Fig. \ref{fig:experiment}(b), is  from a Toptica SHG Pro, stabilized to the desired wavelength using electromagnetically induced transparency (EIT) in an enriched $^{87}$Rb vapor cell~\cite{KP_vapor_EIT_20211}.  The precise wavelength of the 480~nm light is controlled by modifying the wavelength of the 780~nm probe laser using an electro-optical modulator (EOM, NIR-NPX800, Photline Technologies). Shifting the frequency of the 780~nm probe causes a shifting of the EIT condition.  The laser locking circuit compensates for this by adjusting the 480~nm frequency by the same amount.  The output power of the 480~nm light from the ONF is kept constant at 30~$\mu$W. The transmission of the 480~nm light through the ONF is $10\%$ with 300~$\mu$W at the input of the fiber. The 125~$\mu$m fiber supports multiple modes of 480~nm light, all of which, other than the fundamental HE$_{11}$, are lost during the downward taper of the ONF, leading to the low transmission. We maximize the transmission through the ONF by ensuring the light coupled into the input fiber is coupled as much as possible into the fundamental HE$_{11}$ mode.
We use a noise eater (NEL01A/M 425 - 650~nm, Thorlabs) in the path of the 480~nm light to stabilize the power, as, without it, power fluctuations on the order of 5\% would lead to significant fluctuations in the PMT signal. Polarization control (PC), consisting of two quarter waveplates, is used to set the polarization \cite{tkachenko2019polarisation} of the 480~nm field at the ONF, with the optimal condition corresponding to the polarization being quasi-circularly polarized.  In this case, the shape of the evanescent field provides the maximum interaction area for atoms, indicated by maximal loss of atoms from the MOT.
\section{Experimental procedure}\label{sec:methods}
First, we tune the 480~nm laser to select a specific Rydberg state by locking it to a particular wavelength via the EIT locking setup \cite{KP_vapor_EIT_20211}. We compare the wavelength of the 480~nm laser using a wavemeter (HighFinesse WS-6) with the list of frequencies from \cite{PhysRevA.83.052515}.  Unlisted frequencies are calculated using the Alkali-Rydberg Calculator (ARC) package~\cite{SibalicCPC2017}.

\begin{figure}[!t]

    \centering
    \includegraphics[width=0.48\textwidth]{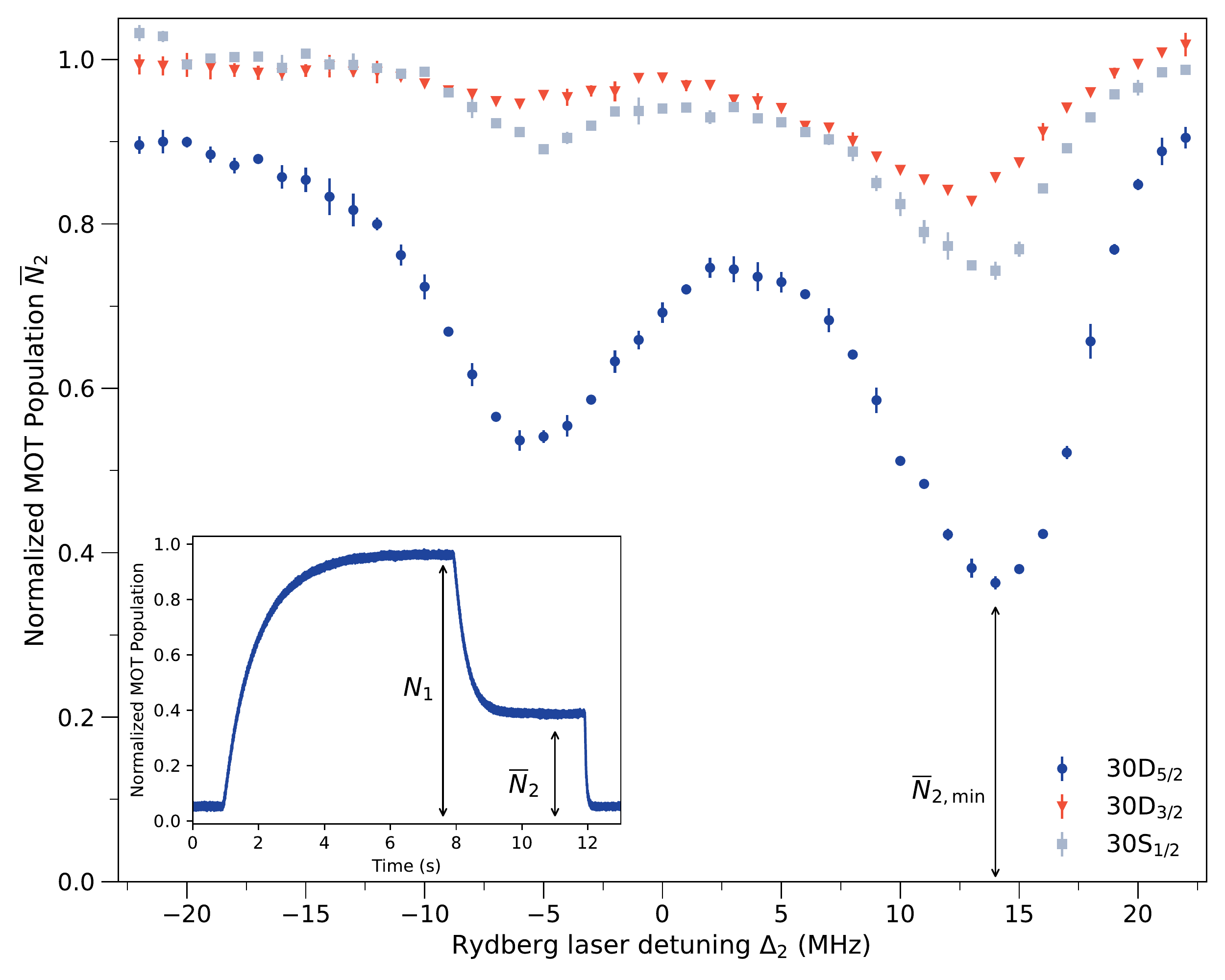}
    \caption{Normalized MOT population, $\overline{N}_2$, after Rydberg excitation as a function of the detuning, $\Delta_2$, of the 480 nm laser for the 5P$_{3/2}$ to 30S$_{1/2}$ (gray squares), 30D$_{3/2}$ (red triangles) and 30D$_{5/2}$ (blue circles) transitions.  The transitions to $n$D$_{5/2}$ states have higher excitation rates than the $n$S$_{1/2}$ and $n$D$_{3/2}$ states due to the larger dipole matrix element. Error bars represent the standard deviation of the 10 experimental measurements of $N_2$ and $\overline{N}_{2,\mathrm{min}}$ represents the minimum observed average population of the MOT after Rydberg excitation. The inset is a sample normalized MOT population curve when $\Delta_2=14$~MHz for the 30D$_{5/2}$ state, indicating how $N_1$ and $N_2$ are determined.}
    \label{fig:results}
    \rule{0.45\textwidth}{0.3pt}
\end{figure}

The experimental sequence (see Fig. \ref{fig:experiment}(c)) consists of two steps. First, atoms are trapped in the MOT for 7 seconds, until  saturation is reached, i.e., the number of atoms in the trap is constant, $N_1$.  We normalize the MOT population in all measurements to this value, i.e., $N_{1}=1$, see inset to Fig. \ref{fig:results}. Second, the 480~nm Rydberg excitation light is switched on, resulting in atoms in the vicinity of the ONF interacting with both the 780~nm light  from the cooling beams and the 480~nm evanescent light field. We keep the 480~nm light on until the normalized atom population in the MOT reaches a new equilibrium, $N_{2} \le 1$, after approximately 4 seconds~\cite{KP_rydberg_generation}. We then determine the dependence of $N_2$ on the detuning, $\Delta_2$, of the 480~nm laser from the two-photon transition,  see Fig. \ref{fig:experiment}(b), by varying $\Delta_2$ from -22~MHz to +22MHz with a step size of 1~MHz and recording the fluorescence signal from the remaining atoms in the MOT using the PMT. The two-photon resonance condition should occur when $\Delta_2=-\Delta_1$, i.e., when the two-photon detuning is zero, ignoring any energy shifts. This gives us an indirect measurement of the rate of Rydberg excitation since any atoms excited to the Rydberg state are lost from the MOT.  We repeat this measurement 10 times for each value of $\Delta_2$ in order to get an average MOT population after Rydberg excitation as a function of the detuning, $\overline{N}_2(\Delta_2)$.  This procedure is then repeated for each specific $n$S or $n$D Rydberg state that we consider. Figure~\ref{fig:results} shows $\overline{N_2}$, as a function of $\Delta_2$ for the $30$D$_{5/2}$, $30$D$_{3/2}$ and $30$S$_{1/2}$ Rydberg states. 
    It is important to note that the excitation to the Rydberg state does not have a defined quantization axis as it occurs within a few hundred nanometers of the center of the MOT where the magnetic field is close to zero. The ONF is positioned so as to overlap with the MOT center. In this region Zeeman shifts of the ground and intermediate states are small (less than 0.5 MHz typically), and depend on the atom position, while the polarization of the photons performing the excitation are also not well defined due to the geometry of the system and the nature of field polarization in the nanofiber's evanescent field ~\cite{fiber_field}. Hence, we can assume that the Zeeman levels are evenly populated. The curves clearly display the two dips resulting from the Autler-Townes splitting of the intermediate level arising from the strong drive of the 780~nm cooling laser, offset from zero due to the non-zero detuning of the cooling laser as the dressed states are asymmetric superpositions of the bare stares. This is the same phenomenon described in the previous work as  coherent and incoherent peaks~\cite{coherent_incoherent,PhysRevA.14.813}.

 The measurements are repeated for the states $n$S$_{1/2}$ for $n\in \left[26,55\right]$, $n$D$_{3/2}$ for $n\in \left[24,65\right]$ and $n$D$_{5/2}$ for $n\in \left[24,68\right]$. Each spectrum (similar to those shown in Fig. \ref{fig:results}) is fit by a skewed Gaussian function to account for the asymmetry arising from the red-shift induced by the AC Stark shift of the 1064~nm laser and the atom-fiber surface interaction.  The fitting function is 
\begin{equation}\label{eq:fitting}
    \begin{split}
    P(\Delta_2)=1-C_1\Phi(A_1(\Delta_2-\mu_1)) \mathrm{Exp}\left[\frac{-(\Delta_2-\mu_1)^2}{2\sigma_1^2}\right]-\\
    C_2\Phi(A_2(\Delta_2-\mu_2)) \mathrm{Exp}\left[\frac{-(\Delta_2-\mu_2)^2}{2\sigma_2^2}\right],
    \end{split}
\end{equation}
where $P(\Delta_2)$ is a fit to $\overline{N}_2$ the normalized final MOT population, $\Phi\left(\mu\right) = 1+\mathrm{erf}\left(\mu\right)$, $\Delta_2$ is the detuning, $C_i$ are the magnitudes of the dips, $\sigma_i$ the widths of the dips, $\mu_i$ the un-skewed Gaussian location, and $A_i$ are the skew parameters. From the fit we extract the mode of the distribution $\nu_2$ which is the position of minimum MOT population.

\begin{figure}[!t]

    \centering
    \includegraphics[width=0.48\textwidth]{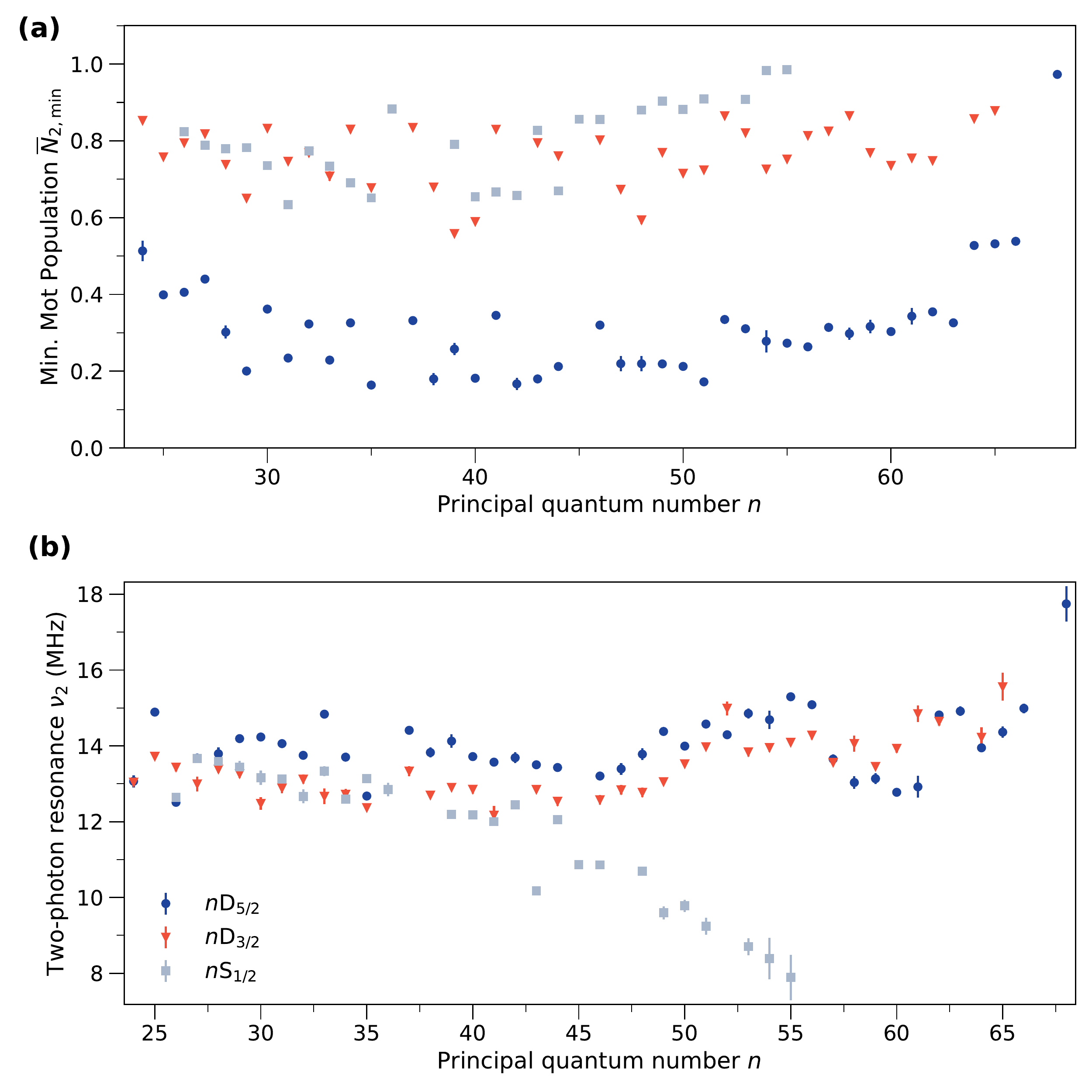}
    \caption{Measured quantities of the $n$S$_{1/2}$ (gray squares), $n$D$_{3/2}$ (red triangles) and $n$D$_{5/2}$ (blue circles) states.
    (a) Dependence of the minimum MOT population $\overline{N}_{2,\mathrm{min}}$ on the principal quantum number $n$.  (b) The position of the two-photon resonance dip $\nu_2$ as a function of the principal quantum number $n$. The $n$S$_{1/2}$ states experience a significant red-shift from the initial $\nu_2=14$~MHz until the excitation stops at $n=55$. The excitation to $n$D$_{5/2}$ and $n$D$_{3/2}$ states experience little energy shift over the range of states measured, with variances of 0.5~MHz and 0.61~MHz respectively. Error bars on the data represent fitting errors of (a)  the dip magnitude and (b) the  dip position.}
    \label{fig:results2}
    \rule{0.45\textwidth}{0.3pt}
\end{figure}

In Fig.~\ref{fig:results2}(a) we plot the minimum observed average MOT population after Rydberg excitation, $\overline{N}_{2,min}$, at the two-photon resonance position for the full range of Rydberg S and D states considered.  Refer to Fig.~\ref{fig:results} to see how $\overline{N}_{2,min}$ is defined for the 30D$_{5/2}$ state as an example. We see that $\overline{N}_{2,min}$ is almost constant regardless of the principal quantum number, $n$, with only the lower and higher $n$ values deviating from the constant loss value. The constant rate of atom loss over a large range of $n$ is because the rate of atom excitation is predominantly dictated by the atom density surrounding the fiber and the overlap with the evanescent field. The rate of loss is dictated by a combination of the excitation and collisional losses of atoms with the Rydberg atoms. This sets an effective maximum excitation rate not determined by the dipole matrix element of the particular Rydberg transition.
These differences at the upper and lower ends could be  due to the drop in power and stability of the 480~nm laser we experience when tuning to longer wavelengths (for lower $n$) and the larger energy shifts and possibility of ionization of Rydberg atoms (at higher $n$).

The two-photon resonance dip position, $\nu_2$, (the position of the minimal MOT population $\overline{N}_{2,min}$) is plotted against the principal quantum number $n$ in Fig.~\ref{fig:results2}(b) and indicates that transitions to the $n$D$_{3/2}$ and $n$D$_{5/2}$ states are less affected by the presence of the ONF than  transitions to $n$S$_{1/2}$ states. Until $n$ approaches 66 there is very little change in the two-photon resonance condition of the $n$D states, indicated by the variances $\sigma_{D_{3/2}}=0.61$~MHz and $\sigma_{D_{5/2}}=0.5$~MHz, while the transition to the $n$S$_{1/2}$ states experiences a clear red-shift before ceasing to be excited at $n\geq 56$. We believe the apparent reduction of the surface interaction of the $n$D states with the ONF surface results from the asymmetry of the electron probability distribution.  The surface interaction with the fiber depends on the orientation of the atomic orbitals with only those aligned parallel with the fiber experiencing a small enough energy shift to be excited, contrasted with the $n$S$_{1/2}$ states which exhibit spherical symmetry. For states $n>60$ the rate of excitation of the $n$D$_{5/2}$ decreases until $n\geq68$ at which point the excitation no longer occurs and the final MOT population is unchanged (see Fig.~\ref{fig:results2}(a)). Speculating that the sudden decrease in the excitation probability arose from ionization of the Rydberg atoms, we tuned the 480~nm laser back to the two-photon resonance condition ($\Delta_2=14$~MHz) for the transition to the $30$D$_{5/2}$ state and repeated the experiment, but observed no excitation signal. Given that the highly excited Rydberg states $\sim{}70$ are known to ionize readily in a MOT~\cite{autoionization}, we believe that at these high $n$ excitations close to the dielectric and in a strong electric field the atoms are ionized, and the ions stick to the surface of the fiber which shields the excitation~\cite{mum_sized_cell}. The impact of DC Stark shifts from stray electric fields or the charging of the ONF are difficult to quantify with no electrodes in the vacuum chamber; however, we assume  these remain relatively constant in our experiments, except for the higher $n$D$_{5/2}$ states when ion deposition on the fiber surface causes enormous shifts. The $n$D$_{5/2}$ states are known to be more sensitive to stray electric fields than $n$S$_{1/2}$ states, so we should expect to see a larger impact on these states if DC Stark effects were significant in these measurements.

To test this hypothesis, we first measured the absorption of a probe laser resonant with the $^{87}$Rb D2 ($F=2\rightarrow F'=3$) transition, obtaining the usual results~\cite{nayak_first_fiber_probe}, indicating there was no issue with the presence of the evanescent field or loss of propagation through the fiber. We then shifted the position of the MOT approximately 0.5~mm along the waist of the ONF using the compensation Helmholtz coils, and performed the Rydberg excitation experiments again. At the new position along the waist the Rydberg experiments worked as expected, supporting our hypothesis that the previous region of fiber had been coated in ions, but was otherwise undamaged.

\section{Numerical Modeling}\label{sec:numerics}
Interactions between atoms and conductive or dielectric surfaces have been investigated thoroughly experimentally~\cite{hinds_lai_schnell_1997, kohlhoff_2016}. However, experiments exploring the interactions between Rydberg state atoms and dielectric surfaces remain limited~\cite{ocola2022control,PhysRevA.97.023418,PhysRevLett.116.133201,mum_sized_cell}. The interactions of Rydberg atoms with dielectric surfaces are expected to be stronger than  for a metal surface, since the dielectric surface can be charged, leading to a resonant interaction between surface polaritons, thereby enhancing the interaction~\cite{mum_sized_cell}. 

For atoms with principal quantum number $n>20$ the quadrupole transitions play a significant role in determining the interaction with surfaces, becoming comparable to the dipole contributions for larger $n$~\cite{StourmPRA2020}.
Interactions with the surface also have an impact on the lifetime of  Rydberg levels, enhancing the spontaneous decay rate in the proximity of a fiber~\cite{Stourm_2019, StourmPRA2020}.

To investigate the importance of these interactions,  we develop a detailed numerical model to compute the rate of Rydberg excitation near an optical nanofiber and quantify the magnitude of the energy shifts.

We calculate the population of the Rydberg state at the end of the experiment by solving the steady-state Maxwell-Bloch equations:

\begin{align}
    \dot{\rho}_{gg} &= \frac{1}{2}\left[\Omega_1\left(\rho_{eg}-\rho_{ge}\right)\right] + \Gamma_1\rho_{ee} + \Gamma_2\rho_{rr} + \gamma_\ell(1-s-\rho_{gg})\nonumber\\
    \dot{\rho}_{ee} &= \frac{1}{2}\left[\Omega_1\left(\rho_{ge}-\rho_{eg}\right)+\Omega_2\left(\rho_{re}-\rho_{er}\right)\right] - \Gamma_1\rho_{ee} + \gamma_\ell\left(s-\rho_{ee}\right)\nonumber\\
    \dot{\rho}_{rr} &= \frac{1}{2}\left[\Omega_2\left(\rho_{er}-\rho_{re}\right)\right] - \Gamma_2\rho_{rr} - \gamma_\ell\rho_{rr}\nonumber\\
    \dot{\rho}_{ge} &= \frac{1}{2}\left[\Omega_1\left(\rho_{ee}-\rho_{gg}\right) - \Omega_2\rho_{gr}+2\Delta_1\rho_{ge}\right] - \left(\frac{\Gamma_1}{2} + \gamma_\ell\right)\rho_{ge}\nonumber\\
    \dot{\rho}_{gr} &= \frac{1}{2}\left[\Omega_1\rho_{er} - \Omega_2\rho_{ge} + 2(\Delta_1+\Delta_2)\rho_{gr}\right] - \left(\frac{\Gamma_2}{2} + \gamma_\ell + \gamma_r\right)\rho_{gr}\nonumber\\
    \dot{\rho}_{er} &= \frac{1}{2}\left[\Omega_2\left(\rho_{rr}-\rho_{ee}\right) + \Omega_1\rho_{gr} +2\Delta_2\rho_{er}\right] - \left(\frac{\Gamma_1}{2} - \frac{\Gamma_2}{2} + \gamma_\ell\right)\rho_{er}
\end{align}

\begin{figure}[!ht]
    \centering
    \includegraphics[width=0.48\textwidth]{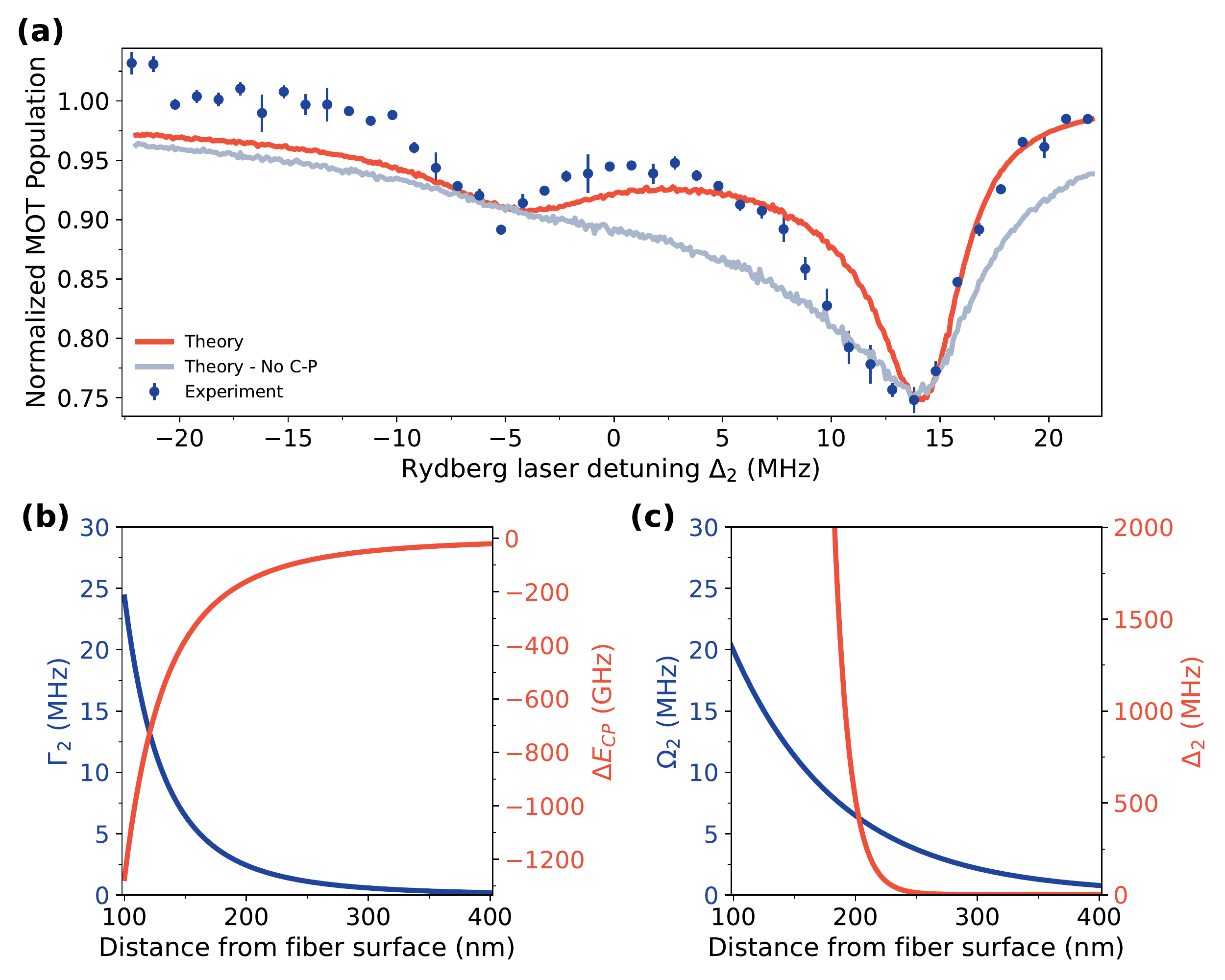}
    \caption{(a) Numerical calculation of the excitation probability of the Rydberg state at steady state for the 30S$_{1/2}$ Rydberg state compared to the experiment. The blue circles show $\overline{N}_2$ from the experiment, the gray line show the numerical results without the Casimir-Polder interaction taken into account, while the red curve includes the Casimir-Polder interaction of the Rydberg atom with the fiber. 
    (b) Numerical calculation of t}he change of the linewidth $\Gamma_2$ of the Rydberg transition (blue) and the Casimir-Polder shift $\Delta E_{CP}$ of the Rydberg energy (red) as a function of distance from the surface for the 30S$_{1/2}$ state. (c) Numerical calculation of the single photon Rabi frequency $\Omega_2$ of the 5P$_{3/2}\rightarrow$30S$_{1/2}$ transition (blue) and the total detuning $\Delta_2$ of the Rydberg state as a function of distance from the fiber surface.
    \label{fig:theory}
    \rule{0.47\textwidth}{0.3pt}
\end{figure}

Here, $\Omega_1$ and $\Omega_2$ are the Rabi frequencies of the 780~nm and 480~nm lasers, respectively, $\Gamma_{1,2}$ the natural linewidths of the transitions, $\Delta_{1,2}$ are the detunings from the respective transitions, $0\leq s \leq 1$ is the mixing ratio describing the relative populations of the ground and intermediate states, and $\gamma_r$ and $\gamma_\ell$ correspond to additional decoherences and losses of atoms from the MOT that are replaced with a new atom in the cooling cycle.  
We calculate the Rydberg excitation probability averaging over 10,000 atoms sampled from various positions of the atomic density near the fiber. For each atom, the Rabi frequencies are calculated from the 480~nm evanescent field and the 780~nm cooling laser intensity. The detunings are calculated including all AC Stark shifts, Doppler shifts from atomic motion, linewidths of the lasers, van der Waals interactions between the atom and the fiber, and the Casimir-Polder energy shifts and lifetime changes of the Rydberg state. In the model, DC Stark shifts are not included as we have no experimental mechanism for quantifying them. They are expected to have minimal impact in comparison to the atom-surface interactions, broadening of the Rydberg transitions, and AC Stark shifts.
The lasers are circularly polarized resulting in a uniform electric field around the fiber, and no specific quantization axis is chosen, as is the case in the experiment.
For the 5S$_{1/2}$ and 5P$_{3/2}$ states the van der Waals potential is approximated by $U_\mathrm{vdW} = A \exp\left[-\alpha(r-a)\right] - C_4/\left(\left(\left|r\right|-a\right)^3\left(\left|r-a\right| + C_4/C_3\right)\right)$ following the treatment in~\cite{PhysRevA.77.042903,PhysRevA.92.013850}. For Rydberg states, the situation is much less trivial~\cite{PhysRevA.105.042817}. Usually only a few low-frequency transitions toward closely lying states substantially contribute to the Casimir-Polder potential~Ref. \cite{EllingsenPRA2011}, and a quasi-static (or zero-frequency) approximation can, therefore, be applied leading to the following form of the potential~\cite{StourmPRA2020}:
\begin{gather}\label{eq:cppotential}
    U_{k} = U_{k}^{(\text{ED})} + U_{k}^{(\text{EQ})}, \nonumber\\
    U^{(\text{ED})}_k = \lim_{\mathbf{r} \to \mathbf{r'}} - \dfrac{1}{2 \varepsilon_0} \sum\limits_{\alpha, \beta} \sum\limits_{k'} d^{(k,k')}_{\alpha} T^{(0)}_{\alpha, \beta}(\mathbf{r}, \mathbf{r'}) d^{(k',k)}_{\beta},
    \nonumber\\
    U^{(\text{EQ})}_k = \lim_{\mathbf{r} \to \mathbf{r'}} - \dfrac{1}{2 \varepsilon_0} \sum\limits_{\alpha, \beta, \gamma, \delta}  \sum\limits_{k'} Q^{(k, k')}_{\alpha, \beta} \dfrac{\partial^2 T^{(0)}_{\beta, \delta}(\mathbf{r}, \mathbf{r'})}{\partial r_\alpha \partial r'_\gamma}  Q^{(k', k)}_{\gamma, \delta}, 
\end{gather}
where we considered only electric dipole (ED) and quadrupole (EQ) allowed transitions from a given state $k$ of interest to $k'$ requiring the dipole ($d^{(k,k')}_{\alpha}$), and quadrupole ($Q^{(k,k')}_{\alpha,\beta}$) operator matrix elements,  $T^{(0)}_{\alpha, \beta}(\mathbf{r}, \mathbf{r'}) = \lim_{\omega \to 0} \frac{\omega^2}{c^2} G_{\alpha, \beta}(\mathbf{r}, \mathbf{r}', \omega)$ with $G_{\alpha, \beta}(\mathbf{r}, \mathbf{r}', \omega)$ being the components of a classical electrodynamical Green's function of the problem.  Note that spatial derivatives here are taken in the Cartesian coordinate system. Under the assumption that atoms occupy each Zeeman sublevel with equal probability, we may write ${\sum\limits_{k'} \dots = \dfrac{1}{2F + 1} \sum\limits_{F', M', M} \dots}$, where $k = (n, F, M), k' = (n', F', M')$ characterize a complete set of quantum numbers for a hyperfine  level. The nanofiber's electrodynamical Green's tensor can be calculated by using the expansion in terms of Vector Cylindrical Harmonics \cite{chew1999waves, KornovanPRB2016}. Finally, the lightshifts can be simply found by $\delta \omega_k = U_k/\hbar$. 

When calculating the emission rate modification due to the presence of the nanofiber, instead of a quasi-static approximation, one has to use the \textit{resonant} approximation as the nonretarded approximation, which is the usual way to find it, is no longer valid:
\begin{gather}\label{eq:linewidth}
    \Gamma_k = \Gamma_k^{(\text{ED})} + \Gamma_k^{(\text{EQ})}, \nonumber\\
    \Gamma^{(\text{ED})}_{k} = \lim_{\mathbf{r} \to \mathbf{r'}} \dfrac{2 \mu_0}{\hbar} \sum\limits_{\alpha, \beta} \sum\limits_{k'} \omega^2_{k,k'} d^{(k,k')}_{\alpha} \text{Im} \left[ G_{\alpha, \beta}(\mathbf{r}, \mathbf{r'}, \omega_{k,k'}) \right] d^{(k',k)}_{\beta} , \nonumber\\
    \Gamma^{(\text{EQ})}_{k} = \lim_{\mathbf{r} \to \mathbf{r'}} \dfrac{2 \mu_0}{\hbar} \sum\limits_{\alpha, \beta, \delta, \gamma} \sum\limits_{k'} \omega_{k,k'}^2 Q^{(k,k')}_{\alpha, \beta} \dfrac{\partial^2 \text{Im} \left[ G_{\beta, \delta}(\mathbf{r}, \mathbf{r'}, \omega_{k,k'}) \right]}{\partial r_{\alpha} \partial r'_\gamma} Q^{(k',k)}_{\gamma,\delta}.
\end{gather}
We compute the change in the lifetime, and hence the linewidths, of the intermediate and Rydberg states when they are in close proximity to the fiber. 

All parameters in the model use the experimentally determined values, aside from the previously discussed Casimir-Polder interaction, van der Waals interaction, and the decoherence terms $\gamma_\ell,\gamma_r$ that are left as free parameters essential in matching the relative height of the Autler-Townes dips. These decoherence terms relate to the rate that Rydberg atoms leave the MOT once excited and the rate at which new atoms enter the MOT - they are difficult to determine experimentally due to the complexity of the dynamics of the MOT. 

The results of the model, omitting the Casimir-Polder interaction, unsurprisingly show large broadening of the transition due to the very high intensity evanescent field close to the fiber surface, as shown in Fig.~\ref{fig:theory}, overcoming the small detuning of the 480~nm laser. Introducing the Casimir-Polder interaction resolves this problem, immediately narrowing the resonant dip to closer match the experiment. The Casimir-Polder shift essentially restricts the excitation to a small region of the evanescent field approximately 250-300~nm from the surface of the fiber, where the Rabi frequency is on the order of a few MHz. Atoms closer than this distance experience such a large energy shift they are no longer resonant with the Rydberg transition. This is shown in Fig.~\ref{fig:theory}(b), and matches the theory of the previous work~\cite{KP_rydberg_generation} where the optimal Rabi frequency was empirically found to be $\sim$2~MHz.
The discrepancy between the model, even with the included Casimir-Polder shift, is most likely due to the fact that it does not include any Rydberg-Rydberg interactions, such as the dipole blockade. Research on the interactions of neighboring Rydberg atoms in the presence of an ONF indicates that it may play a significant role in the excitation dynamics of such a system~\cite{Stourm_2023}. The model also ignores any effects of ionization of the Rydberg atoms, of which the dynamics in a MOT are still an ongoing area of research ~\cite{doi:10.7566/JPSJ.87.054301,Lyon_2017}.

\section{Conclusions}\label{sec:conclusion}
In summary, we have achieved excitation of cold $^{87}$Rb Rydberg atoms next to an ONF via two-photon excitation to a range of $n$S$_{1/2}$, $n$D$_{3/2}$ and $n$D$_{5/2}$ Rydberg states with $n$ ranging from 24 to 68. We have also measured the excitation spectrum, the resonant dip position, and MOT loss rate dependence on the principal quantum number. We observed a strong red shift of the $n$S$_{1/2}$ states that was not detected with the $n$D states. This result suggests that, for future work, the $n$D states, particularly the $n$D$_{5/2}$ states, are the most suitable for ONF-based Rydberg experiments due to their seemingly limited interaction with the fiber that we assume arises from the asymmetry of the electron probability distribution and the reduction of the interaction through the spatial orientation of the atom with respect to the fiber.
We produced a detailed numerical model of the experiment, which includes the Casimir-Polder interaction with the ONF, and confirmed that the Casimir-Polder interaction is of crucial importance in the dynamics of the Rydberg atoms and the dominant energy shift of the Rydberg states in the experiments.
This work provides a critical step forward in the understanding of Rydberg-ONF interactions, which are essential for the continuation of experimental studies of Rydberg atoms at the surface of dielectric, the development of Rydberg-based waveguide QED systems, and the generation of a 1D ordered array of Rydberg atoms for quantum simulations mediated by the nanofiber.

Future experiments will investigate the efficacy of using an optical dipole trap \cite{gupta2022machine} to confine atoms at a fixed distance from the fiber surface, where the Casimir-Polder interaction can be accurately compared to calculations and the loss dynamics are significantly easier to determine without having to consider the full MOT system.  This could also be experimentally favorable by reducing the ionization of highly excited Rydberg states caused by the atom-surface interactions. Additionally, adapting the conventional two-color dipole trap~\cite{PhysRevA.70.063403,PhysRevLett.104.203603,PhysRevLett.113.263603} to allow for the trapping of Rydberg atoms, potentially through the use of magic wavelengths~\cite{Bai_2020} would be a significant step forward in the development of optical nanofiber and Rydberg atom based quantum devices, enabling the investigation of any effect the presence of the ONF has on the Rydberg blockade~\cite{Stourm_2023}.

\section*{Appendix: Technical Details}\label{appendix}

\subsection{Theoretical Model}

\begin{figure}[!ht]
    \centering
    \includegraphics[width=0.45\textwidth]{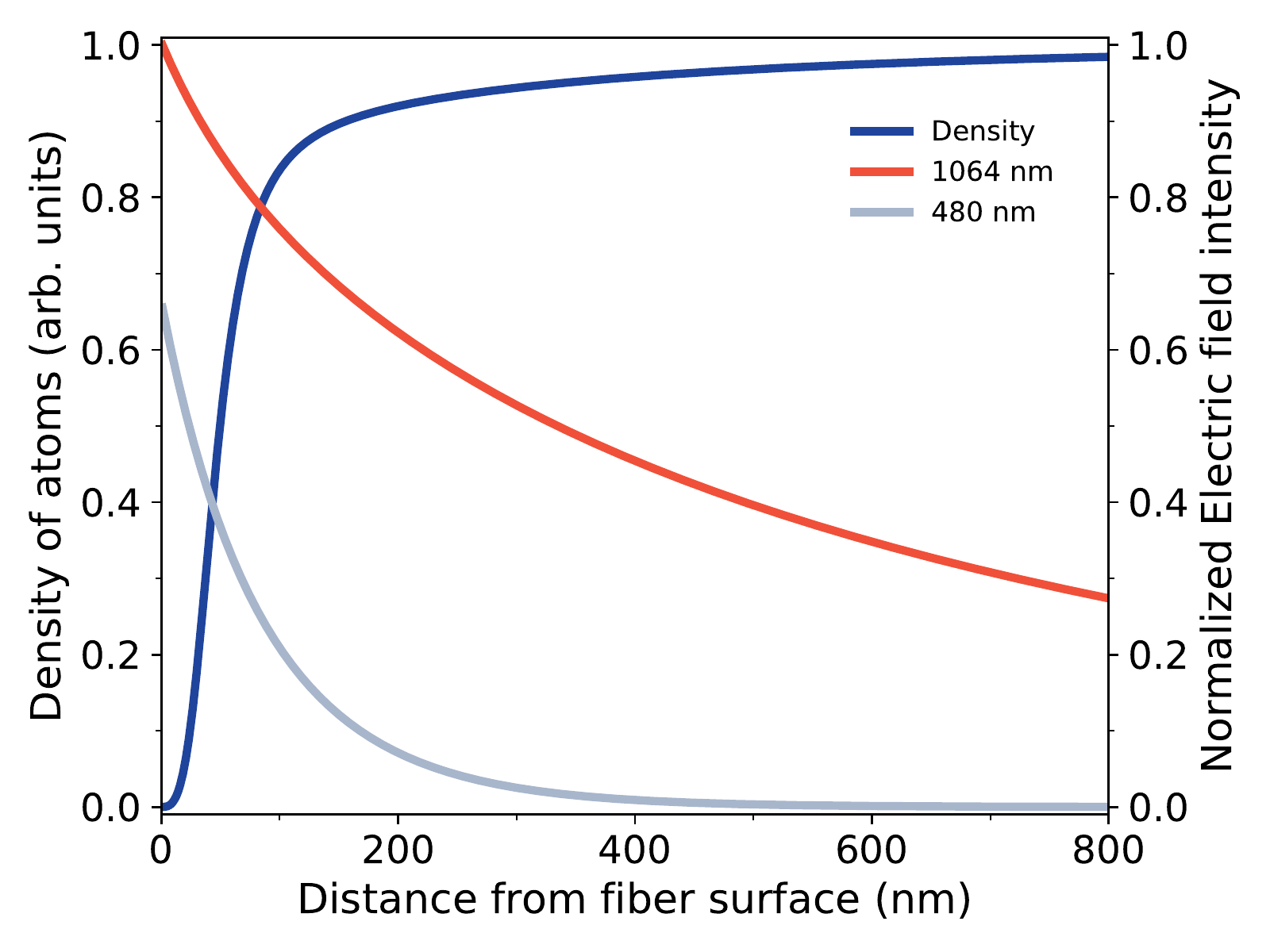}
    \caption{Normalized density of the atoms (blue) near the ONF surface due to the van der Waals forces and attractive force of the 1064~nm laser in the fiber. The evanescent fields of the 1064~nm (red) and 480~nm (gray) lasers are also shown, normalized to the highest intensity of the 1064~nm light.}
    \label{fig:density}
\end{figure}
The evanescent fields of both the 1064~nm and 480~nm light at the waist of the ONF are calculated explicitly following~\cite{PhysRevA.96.023835} with the polarizations set to be quasi-circular due to the non-zero longitudinal component of the electric field in the fiber.  The optical power is set to experimental values. The Rabi frequencies, $\Omega$, and AC stark shifts, $U_{AC}$,   are directly calculated from the intensities  as
\begin{gather}
    \Omega = \frac{-\mathbf{d}\cdot|\mathbf{E}|}{\hbar}\\
    \nonumber U_\mathrm{AC} = \frac{1}{4}|E|^2\left[\alpha^{(0)}\left(\omega\right) - i\alpha^{(1)}\left(\omega\right)\frac{\left(\mathbf{u}^*\times \mathbf{u}\right)\cdot \mathbf{F}}{2F}\right.\\\left. - \alpha^{(2)}\frac{ 3\left[(\mathbf{u}^*\cdot\mathbf{F})(\mathbf{u}\cdot\mathbf{F})+(\mathbf{u}\cdot\mathbf{F})(\mathbf{u}^*\cdot\mathbf{F})\right]-2\mathbf{F}^2}{2F(2F-1)}\right]
\end{gather}
where $\mathbf{d}$ is the dipole operator, $\alpha^{(0)}\left(\omega\right),\alpha^{(1)}\left(\omega\right),\alpha^{(2)}\left(\omega\right)$ are the scalar, vector, and tensor polarizabilites of the atom, $E$ the electric field and $\mathbf{F}$ the total angular momentum~\cite{LeKien2013}. 

The explicit forms of the dipole and quadrupole matrix elements of Eq.~\ref{eq:cppotential} are given by \cite{KienEPJD2013, KienPRA2018}:
\begin{gather}
  \langle n'F'M' | \mathbf{d} | n F M \rangle =  (-1)^{J' + I' + F + 1} 
    \sqrt{(2F'+1)(2F+1)} \\ \times \begin{Bmatrix} F' & 1 & F \\ J & I & J'
    \end{Bmatrix} \langle n' J' || \mathbf{d} || n J \rangle,\nonumber\\
    \langle n' F' M' | Q_{\alpha, \beta} | n F M \rangle = 3 e u_{\alpha \beta}^{(M'-M)} (-1)^{F'-M'} \\ \times \begin{pmatrix} F' & 2 & F \\  -M' & M'-M & M \end{pmatrix} \langle n'F' || T^{(2)} || nF \rangle,\nonumber
\end{gather}

\noindent where $I$ is the nuclear spin, $J = L + S$ is the total electron momentum, $\langle n' J' || \mathbf{d} || n J \rangle$, is the reduced dipole matrix element, the quantity in curly brackets is a $6j$-symbol, while in round ones - $3j$-symbol, $e$ is the electron charge, matrices $u^{(q)}_{\alpha, \beta}$ with $q=\{ -2,-1,0,1,2 \}$ written in Cartesian coordinates can be found in Ref. \cite{KienPRA2018}, and $\langle n' F' || T^{(2)} || n F \rangle$ is the reduced matrix element of a tensor operator $T^{(2)}_q = 2 \sqrt{\dfrac{2 \pi}{15}} R^2 Y_{2,q}(\theta,\phi)$. For a quadrupole operator matrix element, one can decouple the angular and radial parts. The latter, as well as the reduced dipole matrix element, can be found, for example, by using the ARC package \cite{SibalicCPC2017}. 

The normalized atomic density surrounding the fiber is shown in Fig.~\ref{fig:density}, and is calculated following~\cite{grover-phdthesis} from the expression
\begin{equation}
    \rho=\Theta\left(r-a\right) 1/\left[1-\frac{U_\mathrm{vdW}+U_\mathrm{AC}}{\frac{3}{2} k_\mathrm{B} T}\right]
\end{equation}
where $U_\mathrm{vdW}$ is the van der Waals potential of the fiber, $U_\mathrm{AC}$ is the potential formed from the AC Stark shift of the 1064~nm laser, $\Theta\left(r-a\right)$ is the Heaviside function fixing the potential outside the surface of the fiber of radius, $a$, and $T$ is the average temperature of the atoms in the MOT. The density is assumed to tend towards a constant value away from the surface of the fiber. The competition between the kinetic energy of the atoms and the attractive forces near the fiber causes atoms to accelerate in the vicinity of the ONF, leading to a drop in density close to the fiber surface. This calculation gives a result that closely matches that presented in previous work~\cite{PhysRevA.77.042903}, minus the oscillations present in the full quantum calculation.


\begin{acknowledgments}
The authors would like to thank P Cheinet, F Le Kien, S Lepoutre and K. Mølmer for very insightful discussions, K Karlsson and M Ozer for technical assistance, Y Saleh for contributions to data collection, the Nanofabrication Section for Scanning Electron Microscope access, and Mechanical Engineering Section of the Research Support Division at Okinawa Institute of Science and Technology Graduate University. \\
	
This research is supported by Okinawa  Institute  of  Science  and  Technology  Graduate  University, 
Okinawa, Japan and Japan Society for the Promotion of Science (JSPS) Grant-in-Aid for Scientific 
Research  (C)  under  Grants  No.  19K05316  and  No.  20K03795  and  Grant-in-Aid  for  Scientific 
Research (Early Career) No. 22K13986.  Investments for the Future from LabEx PALM (ANR-10-LABX-0039-PALM). Danish National Research Foundation through the Center of Excellence “CCQ” (Grant agreement no.: DNRF156). 
\end{acknowledgments}



\bibliography{bibliography}

\end{document}